# "MS-Patch-Clamp" or the Possibility of Mass Spectrometry Hybridization with Patch-Clamp Setups for Single Cell Metabolomics and Channelomics

## Oleg Gradov[1,*], Margaret Gradova[2]

[1]Talrose Institute for Energy Problems of Chemical Physics, RAS, Moscow, Russia
[2]Semenov Institute of Chemical Physics, RAS, Moscow, Russia

### Email address:
gradov@center.chph.ras.ru (O. Gradov)

**Abstract**: In this projecting work we propose a mass spectrometric patch-clamp equipment with the capillary performing both a local potential registration at the cell membrane and the analyte suction simultaneously. This paper provides a current literature analysis comparing the possibilities of the novel approach proposed with the known methods, such as scanning patch-clamp, scanning ion conductance microscopy, patch clamp based on scanning probe microscopy technology, quantitative subcellular secondary ion mass spectrometry or "ion microscopy", live single-cell mass spectrometry, in situ cell-by-cell imaging, single-cell video-mass spectrometry, etc. We also consider the ways to improve the informativeness of these methods and particularly emphasize the trend at the increasing of the analysis complexity. We propose here the way to improve the efficiency of the cell trapping to the capillary during MS-path-clamp, as well as to provide laser surface ionization using laser trapping and tweezing of cells with the laser beam transmitted through the capillary as a waveguide. It is also possible to combine the above system with the microcolumn separation system or capillary electrophoresis as an optional direction of further development of the complex of analytical techniques emerging from the MS variation of patch-clamp.

**Keywords**: patch-clamp, chanellomics, chanellome, scanning ion conductance microscopy, *in situ* cell-by-cell imaging, quantitative subcellular secondary ion mass spectrometry, single-cell video-mass spectrometry, ion microscopy, STM patch-clamp, membrane receptors, single-channel electrophysiology.

## 1. Introduction

Among numerous specialized methods of the fine single cell analysis based on microelectrode and microcapillary techniques, such as extracellular and intracellular voltammetry, capillary electrophoresis and microcolumn separation, etc. [1], the most common one is the method of local potential registration of single channels also known as patch-clamp. It was first developed in the middle 1970-th [2], became widely available at the early 1980-th [3] and now patch-clamp appeared to be a multifunctional precision quantitative method for elecrophysiological registration of activity and reactivity of both single ion channels and their complexes functionally related according to the cytophysiological criteria. A complex of channels in the framework of the above approach is called a "channelome" [4,5], and the corresponding branch of science is known as "chanellomics" [6], which studies the forms of either electrophysiological or electrobiochemical response of the ion channel system, as well as their biophysical dependencies.

However, providing registration of a single ion channel activity, patch-clamp does not allow to perform chemical analysis. Thereby, since early 1990-th the attempts were made to combine local potential registration techniques with various analytical methods, in particular, with capillary electrophoresis [7-10], but this method is obviously not exhaustive, since there are fundamental limitations on combining capillary electrophoresis with electrochemical detection [11]. Thus, only assuming the nature of the substance detected it is possible to obtain a qualitative result. Patch-clamp combined with electroporation [12] and respectively with a single cell PCR [13] combined with electrodiffusion transport activity during patch-clamp [14] is well adapted for membranology and molecular genetics of single cells, such as neurons [15,16], but is not intended for the analysis of other cell constituents. The authors of a pioneer work [15] describe this as follows: "combination of patch-clamp and molecular biology techniques has made it possible to characterize the pharmacological and biophysical properties of ion channels in single neurons and to screen for expression of specific mRNAs in the same cell".

## 2. Ways of Chemical Registration.

A palliative solution (with respect to the existing techniques) is the combination of scanning patch-clamp methods with ionic conductance microscopy of living cells [17][1], also known as scanning patch-clamp [18][2]. This method has been successfully hybridized with the latest scanning techniques of the mass spectroscopic tissue imaging using a scanning microcapillary [19].

In contrast to the electrophoretic patch-clamp detection variation, mass spectrometry allows registration and distinction between various chemical cell constituents. Since 1980-th, when imaging secondary ion mass spectrometry, also known as ion microscopy, provided quantitative concentration determination of boron, calcium, magnesium, potassium, and sodium in various parts of the living cell [20,21], and later with some of these elements, in particular, with boron [22], an isotopic analysis was performed – a quantitative subcellular secondary ion mass spectrometry (SIMS) imaging, this area has become a development trend of biological MS imaging. It is quite obvious that the possibility of inorganic ion detection in this way allows one to perform temporal registration and imaging of ion channel activity and mapping of the receptors which control the influx of inorganic ions [23]. This correlates well with the possibility of inorganic ion imaging during patch-clamp and patch-clamp recording with temporal resolution, e.g., chloride imaging and cell-attached patch-clamp recordings for the study of how a chloride efflux inhibits single calcium channel open probability[3] [24].

It is of particular interest that, in addition to the inorganic ion content analysis, for example, calcium stores in cells [25], MS imaging also allows to determine the content of cell surfactants, for example, cholesterol in the membranes of individual cells [26], that provides data interpretation at the physico-chemical level. This is even more promising in terms of dynamic secondary ion mass spectrometry analysis [27], since in this case the comparability with the temporal patch-clamp registration can be clearly revealed.

To date, the main principles of live single-cell mass spectrometry have been developed [28], capable of operating *in vivo* at atmospheric pressure [29] and detecting even midget fragments [30]. They are used for various tasks at different organization levels of living matter, from single-cell metabolomics [31] to single-cell ecophysiology, often in combination with other microspectral, in particular, optical methods [32]. As well as usual patch-clamp, which in its planar patch clamp configuration [33-35][4] is capable of screening at large tissue layers, the above mass spectrometric methods also allow working with large tissue layers [36,37] (up to the so-called "leaf spray" method for direct chemical analysis of living plants [38]) and cell populations, performing *in situ* cell-by-cell imaging with cellular resolution [39,40].

Thus, it is possible to study precisely single cell activity and intercellular interactions in the populations and tissues. Unfortunately, there have been no attempts to apply any of these methods to the study of electrobiochemical activity and the chemistry of electrophysiological regimes in channelome, although this follows directly from the above considerations. However, its implementation requires combination of the input capillary of the detection / registration device with the patch-clamp capillary, which is impossible in some of the similar methods, since they are based on the imaging principles different from the required ones. Thus, it is almost meaningless to combine patch-clamp with MALDI imaging without using the capillary also as a laser waveguide, operating simultaneously as a tool for laser trapping and tweezing. It also does not make technical sense to combine patch-clamp with any other detection methods which do not use the capillary and the suction.

The novel methods for single-cell video-mass spectrometry, combining the optical and mass spectrometric methods of registration and imaging [41-43], suggest the comparison with the combination of patch-clamp and optical methods [44], that is often used in analyzing of electrophysiological and signaling activity [45], in particular, during single channel registration [46], especially in its fluorescence variation [47] special attention should be paid to the possibility of the laser beam passing through the capillary as a waveguide, which enables a synchronized laser ionization and combination of the optical tweezers with patch clamp for the study of cell membrane electromechanics [48], that was previously performed using standard patch-clamp methods [49-51].

---

[1] See also patents: "Scanning ion conductance microscopy for the investigation of living cells" (WO 2008015428 A1; EP 2047231 A1; US 8006316 B2), based on the fundamental technique, patented as "Scanning ion conductance microscopy" (WO 2009095720 A2, US 20110131690 A1, EP 2238428 A2).

[2] See patents: "High resolution patch clamp based on scanning probe microscopy technology and operating method thereof" (CN 102071135 A) and "High resolution patch clamp device based on scanning probe microscopy technology" (CN 201654064 U).

[3] It is worth  mentioning that SIMS is developed for the work with the frozen tissues rather than with native ones.

[4] See also general patent "Planar patch clamp electrodes" (US 6,999,697) and its patent application (US 2004/0168912 A1).

However, both laser trapping and patch clamp are characterized by the forces, providing retraction of the cell (or its area, as it occurs in the outside-out method), while standard methods of MS-imaging do not use retraction. Hence, the problem of combining the patch-clamp principles with the suction into a mass spectrometer is of great scientific relevance and novelty, and it can be successfully solved both for patch clamp in loose contact (suitable for potential registration from the whole cell or its membrane at the area of three orders of magnitude higher than in single channel registration - up to 200 $\mu m^2$), and for conveyor and automated the whole-cell method.

## 3. Simple Technical Problems to be Solved

Thus, the main task is to design a head with a capillary and its connection to the MS-skimmer in such a way that the chemical signal from the cell could be registered by the spectrometer. From this perspective the most effective is FT-MS registration, where such contacts are rather common, [52,53], and FTICR-MS itself is known to be the most effective tool for the analysis of macromolecular, and especially biomacromolecular, samples, among the existing MS-methods.

Despite the applicability of the capillary variation in polysaccharide analysis using capillary electrophoresis with mass spectrometric detection [54] and in capillary liquid chromatography with electrospray mass spectrometry in biochemical analysis [55], they can not be considered more effective in registration during the patch-clamp procedure, although we can not exclude the perspectives of ultramicrocolumn and, in fact, submicroscopic and nanoscale separation for detection during patch clamp experiments in the future.

It is well known that in gas mass spectrometry the question about the comparison of the capillary and skimmer interfaces efficiency still remains controversial [56], but in FT-MS and related techniques the effect of capillary – skimmer potential difference possesses a complex value for detection efficiency [57]. It should be noted that the resulting mass spectrometric patch-clamp will represent in fact not only a single patch-clamp technique as it is known now, since FTICR-MS allows registration not only of the ionic channel activity, but also of the chemical composition of the medium, including macromolecular compounds.

Moreover, FTICR is applicable for chiral supramolecular structures measurements [58], as well as for the study of asymmetric reactions [59]. This method is compatible with ESI-MS – electrospray ionization mass spectrometry [60], which has already provided significant results for neurobiochemistry [61], and hence, with nanospray desorption – electrospray ionization mass spectrometry, useful in cell-by-cell imaging and tissue imaging [19]. Adaptation of the existing equipment for microdoses is, in fact, a physico-technical problem, which seems to be rather feasible. Since it is not closely related to the biochemistry of the process, it will be considered elsewhere.

## 4. Conclusions

The idea of "MS-patch-clamp" in its current state is not simply a research method, but a fundamental approach that can be extended to a large number of detection, ionization and desorption methods and various ways of the analyte supplying. Both high resolution and high sensitivity is required, though the orbitrap exceeds FTICR-MS in sensitivity. All of these instruments provide a high mass accuracy (<2-3 ppm with external calibrant and <1-2 ppm with internal), a high resolving power (up to 240,000 at m/z 400), a high dynamic range and high sensitivity [62,63]. FTICR может provide a detection limit of approximately 30 zmol ($\approx$ 18 000 molecules) for proteins with molecular weights ranging from 8 to 20 kDa [64]. To date the detection limit for accelerator mass-spectrometry in different analytical protocols is not as low as femtomolar [65,66] and attomolar [67,68], but even subattomolar [69] and zeptomolar [70-72] concentrations, corresponding to the sensitivity at the atomic level and satisfying the trend to the «accelerator mass spectrometry-isotope measurements at the atomic level» [72]. Thus, the accuracy of MS-patch-clamp analysis can be significantly improved with the implementation of «accelerator MS – patch-clamp» concept.

A pilot volunteer project in this area is now starting at RAS Center of Mass spectrometry, IBCP RAS and IEPCP RAS, but it is also technically possible to design similar instrumentation based on more simple or even alternative principles as a DIY project [73], since many novel ideas and findings during patch-clamp experiments traditionally have been obtained using a combinatorial DIY approach [74-76] which provided a lot of new data for cellular and molecular biology.

## Acknowledgments

We thank our colleagues from the Laboratory of Ion and Molecular Physics (INEPCP RAS) and also from the Laboratory of Molecular and Ion Physics (MIPT) for the helpful comments on the manuscript.

We thank the whole staff of the Department of Metrology and Measurement Techniques in the Institute of Geochemistry and Analytical Chemistry (GEOKHI RAS) for providing all the details and tools needed for constructing of the early hybridized patch-clamp setup applicable for "MS patch-clamp".